# Synaptic MIS silicon nitride resistance switching memory cells on SOI substrate

A.E. Mavropoulis, N. Vasileiadis, P. Normand, V. Ioannou-Sougleridis, K. Tsakalos, G. Ch. Sirakoulis, and P. Dimitrakis

*Abstract*— In this work, the comparison of resistive memory MIS single-cells without selector (1R), having silicon nitride as switching dielectric, fabricated on SOI and bulk Si wafers is demonstrated. Comprehensive experimental investigations reveal the advantage of SOI substrate over bulk Si in terms of cycling endurance, cell-to-cell variability, retention and data loss rate.

## I. INTRODUCTION

Silicon-on-insulator (SOI) substrates are of special importance for nanoscale CMOS, photonic and RF silicon technology [1]. Meanwhile, the back Si interface with BOX add extra functionalities in CMOS devices like easy back-gate operation and resistivity control of handle wafer [2]. Nevertheless, a functional SOI device design should take into account several new phenomena that appeared such as floating-body, parasitic bipolar transistor, BOX charge polarization etc. [3, 4] Volatile and nonvolatile memories exhibit enhanced performance also taking advantage of the radiation tolerance [5], low-power [6, 7] and silicon channel isolation through the buried oxide (BOX) [8]. In addition, SOI is an ideal substrate for testing new and emerging memory concepts like ferroelectric capacitance memories [9, 10].

Resistive random-access memory (RRAM) technology is considered as the most promising candidate to replace Flash NVMs and realize the storage class memory concept [11, 12]. In addition, it has been explored for low-power circuits for neuromorphic and in-memory computing systems [13, 14] as well as edge-computing for internet of things applications [15, 16]. The demonstration of functional RRAM integration in 28nm FDSOI technology has been successfully demonstrated [17, 18]. Also, the radiation-tolerance of SOI-RRAM has been investigated [19] suggesting their premium functionality in harsh environments with strict power and reliability constraints. Recently, fabrication of resistive memory cells on SOI substrate has been utilized to resolve the influence of parasitic effects in memory cell operation [20, 21].

In this work, the exploration of silicon nitride resistive memory single-cells (1R) operation characteristics and performance are presented. The report is organized as follows: Section II refers to the fabrication and measurement experimental details. Next, the experimental results and the related discussion are following. In the final section, the main conclusions of our investigations are summarized.

## II. EXPERIMENTAL

### A. Device fabrication

We started with a Smart-Cut SOI substrate having 100nm *p*-type Si (50 μΩ·m) superficial layer and 200nm BOX layer. The SOI was implanted with Phosphorus ions ($1\times10^{15}$ cm$^{-2}$ at 40keV) through a 20nm $SiO_2$ sacrificial layer with a subsequent rapid thermal annealing at 1050°C for 20s. The role of this oxide layer is three-fold. First, it is used to adjust the peak of the implantation profile of dopants and second to prevent the surface damage of SiN layer. Additionally, it is used to mitigate the out-diffusion of P atoms during activation annealing. The final doping of the Si layer was about $5\times10^{19}$ cm$^{-3}$ [20] allowing its use as bottom electrode (BE) of the memory cell. Following, a 7 nm $SiN_x$ layer was deposited by LPCVD at 810°C, using ammonia ($NH_3$) and dichlorosilane ($SiCl_2H_2$) gas precursors. Finally, 100μm × 100μm top-electrodes (TE) were defined by photolithography and metal lift-off. Metallization comprises a sputtered 30nm Cu layer covered by 30nm Pt to prevent oxidation of Cu. Additional etching and photolithography steps were employed in order to form the BE contact depositing 100nm thick Al metal electrode. For the sake of comparison, the same cell structures were fabricated on bulk n$^{++}$-Si wafer. The final MIS structure of the memory single cells is shown in Figure 1 [21].

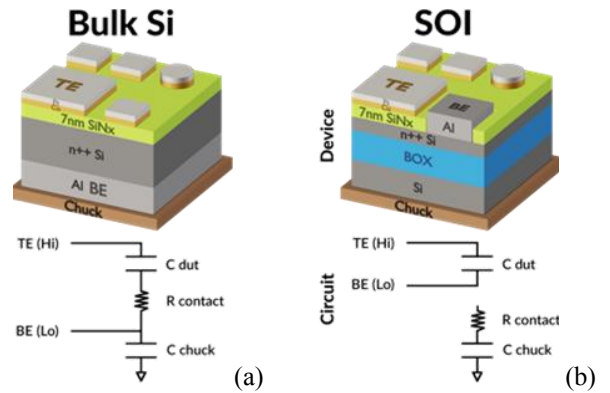

Figure 1. MIS 1R RRAM cells on (a) bulk and (b) SOI substrates. (Inset) The device is isolated from prober's chuck in the case of SOI.

Research supported by HFRI through "LIMA-chip" project (Proj.No. 2748) and by GSRT-Operational Programme NSRF 2014-2021 "3D-TOPOS" (MIS 5131411). A.E. Mavropoulis, N. Vasileiadis, P. Normand and P. Dimitrakis are with the Institute of Nanoscience and Nanotechnology, NCSR "Demokritos", Aghia Paraskevi 15341, Greece. P. Dimitrakis is the corresponding author, p.dimitrakis@inn.demokritos.gr. N.Vasileiadis, K. Tsakalos, I. Karafyllidis and G.Ch. Sirakoulis are with the Department of Electrical and Computer Engineering, Democritus University of Thrace, Xanthi 67100, Greece.

## B. Experimental and Measurement procedures

The DC I-V characteristics of the fabricated devices were measured using the HP4155A and Tektronix 4200A equipped with fast current preamplifiers. Impedance spectroscopy measurements in the range 100Hz – 1MHz were also performed on devices either in high/low-resistance state (HRS / LRS) using the HP4284A Precision LCR Meter and 0.1Hz – 500kHz MFIA Zurich Instruments. All measurements were performed on a Cascade Summit 12000 semi-automatic wafer prober at room temperature in ambient air (RH=45%). A special measurement setup and software were used to realize smart pulse tuning protocol, in order to achieve multi-resistance states as well as analog performance of the examined memristive devices, that is presented elsewhere [22, 23]. Random telegraph noise (RTN) measurements were performed using a digital oscilloscope and a battery-powered SR570 I/V converter for current recording under a constant read voltage, typically 0.1V, for 100s. All experiments were performed keeping BE grounded and applying the bias on TE.

## I. RESULTS AND DISCUSSION

### A. DC operation and Device Characteristics

Figure 2(a) depicts a large number of static I-V characteristics for the examined SOI resistive memory cells using current compliance $I_{CC}$ =100μA. Obviously, during the first voltage sweep a forming process is taking place, i.e., the intense Joule heating and high electric field, cause sometimes soft breakdown. According to experimental data and literature [24, 20] the most probable model explaining the resistance switching in SiN$_x$ thin layers is due to the migration of nitrogen anions, N$^-$, towards the positively biased TE creating mobile nitrogen vacancies, $V_N^+$, and silicon dangling bonds. The anions are moving through the native defects of Si and become neutral when they reach the TE metal. Thus, conductive paths (or filaments, CF), made of $V_N^+$ are formed [28] and when the filaments connect TE and BE, switching from HRS to LRS occurs. The application of inverse polarity voltage on TE push accumulated nitrogen ions, at the SiN$_x$/TE interface, back to the bulk causing the partial dissolution of the CF. The process is similar to the oxide-based resistive memory cells and other Valence-Change Memory (VCM) devices [25-27].

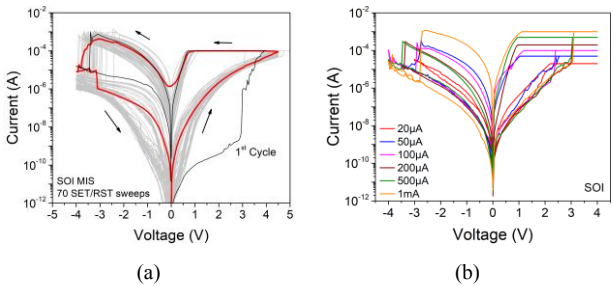

Figure 2. (a) A plethora of successive I-V round sweeps under constant $I_{CC}$=100μA (b) Typical I-V characteristics for different $I_{CC}$ values, denoting the multi-state operation.

Figure 2(b) presents multiple current-voltage measurements under different current compliance, $I_{CC}$ values. Obviously, the fabricated memory cells can be programed (SET) at various resistance levels according to the specific $I_{CC}$ value. Additionally, the abrupt and almost symmetric switching characteristics are evident. The higher the $I_{CC}$ value the higher the number of the CFs that are formed, or equivalently the thicker the diameter of a single filament. Furthermore, the measurements shown in Fig.2(b) suggest that the devices exhibited current self-compliance, i.e., the current before RESET, during the negative voltage sweep, is similar to $I_{CC}$ value, which is higher in for bulk devices [20]. This is mainly attributed in bulk Si to the parasitic capacitance of the chuck in the probe station (Fig.1) and the slightly higher resistance of the BE, in SOI. The statistics of $V_{SET}$ and $V_{RESET}$, where the mean values for SOI devices are greater compared to bulk ones. However, the variation of $V_{RESET}$ is more uniform in SOI than bulk devices, while $V_{SET}$ uniformity if the same.

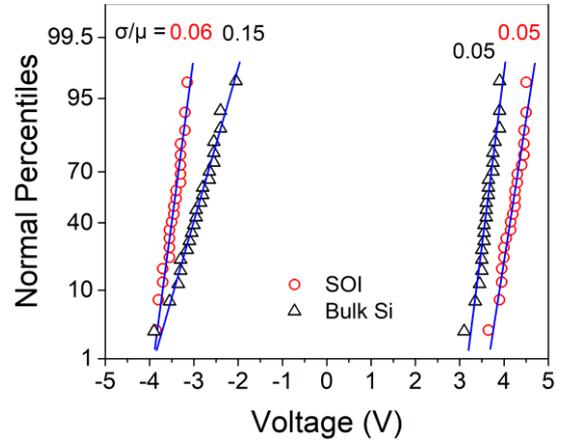

Figure 3. Statistical analysis of the measurements in Fig.2(a) for $V_{SET}$ and $V_{RESET}$.

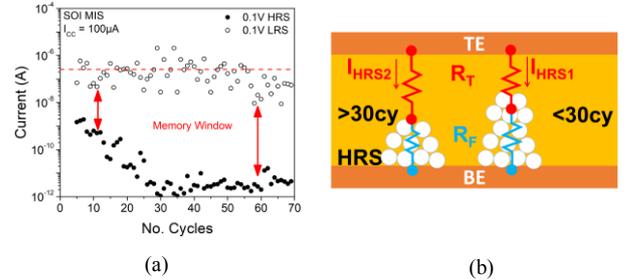

Figure 4. (a) The evolution of memory window over cycling. (b) CF size in the case of <30 cycles and >30 cycles.

Further analysis of measurements in Fig.2(a) reveals that the memory window of the SOI cells is increasing after consecutive I-V cycling and is stabilized after 30 cycles, as shown in Fig. 4(a). This is mainly attributed to the length of the dissolved CF(s). As shown in Fig.4(b) the total resistance of a memory cell at HRS should be approximated by [29]

$$R_{HRS} = R_T + R_F \quad (1)$$

where $R_T$ and $R_F$ represent the resistance of the dissolved and remaining part of the CF respectively. In case the device is cycled by less than 30 times, $R_T$ increases and $R_F$ decreases. $R_T$ and $R_F$ remain constant when the number of cycles exceeds 30. The relation between the length of the dissolved and HRS is empirically modeled by [29]

$$R_{HRS} = \rho_{Si,CF} \frac{t_{SiN}}{S} \left(e^{\frac{x}{\kappa}} - \frac{x}{t_{SiN}}\right) = R_{LRS} \left(e^{\frac{x}{\kappa}} - \frac{x}{t_{SiN}}\right) \quad (2)$$

where $R_{LRS}$ is the resistance of the CF assuming that it is a cylindrical Si nanowire with length equal to the thickness of $SiN_x$ layer and $S$ is the cross-section area, $x$ is the length of the dissolved CF and $\kappa$ is an empirical parameter. It should be emphasized that this behavior of the memory window is not observed when the memory cell was SET using incremental step pulse programming (ISPP) [20]. Hence, the observed evolution of the memory windows should be related to the initial forming process.

### B. Constant voltage stress measurements

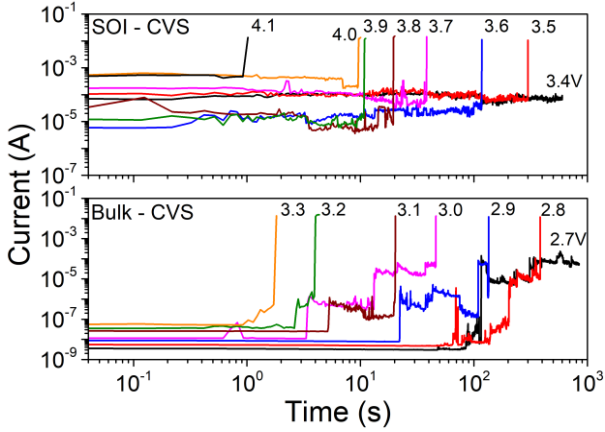

Figure 5. Constant voltage stress (CVS) test : Leakage current vs time for various stress voltages until breakdown occurs.

In figure 5, the constant-voltage-stress measurements are presented. Evidently, the SOI memory cells are broken-down when the stress voltage, $V_{stress}$, becomes higher than 3.5V, while breakdown of cells in bulk wafer starts at 2.8V. This is because of the higher resistance of the SOI BE, as mentioned previously. Furthermore, in the case of memory cells on bulk Si wafer the *I-t* plots under CVS present a quantized breakdown mechanism, which is not detected in CVS measurements of SOI memory cells. This finding as well as the breakdown mechanism is under investigation.

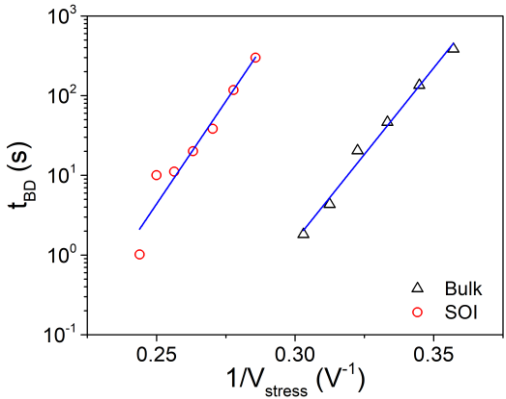

Figure 6. Time-to-breakdown during CVS as a function of the reciprocal stress voltage.

Preliminary results from the analysis of the measurements shown in Fig.5 revealed that the time-to-breakdown $t_{BD}$ is proportional to $exp(1/V_{stress})$, as shown in Fig.6, for both SOI and bulk memory cells. Previous experimental studies of $Cu/SiN_x$ (ca. 50nm) CVS measurements refer excellent Cu diffusion resistance and $t_{BD}$ depends on the current conduction mechanism, i.e., Poole-Frenkel, Fowler-Nordheim etc [30-31].

### C. Impedance spectroscopy measurements

Typical experimental Nyquist plots for memory cells on SOI at LRS (SET) are shown in Fig. 7. Obviously, these data can be modeled by a parallel resistor-capacitor ($R_p||C_p$) equivalent circuit and a series resistance ($R_s$) due to cable connections etc. The impedance Z of this circuit is described by

$$Z = Z' - jZ'' = \frac{R}{1+\omega R^2 C^2} - j\frac{\omega R^2 C}{1+\omega R^2 C^2} \qquad (3)$$

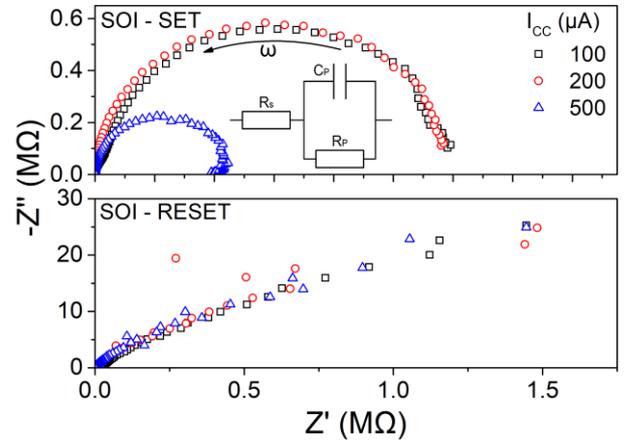

Figure 7. Nyquist plots of the examined 1R MIS RRAM cells at LRS and HRS. (inset) Corrsponding $R_p||C_p$ equivalent circuit to fit the experimental data.

Contrary, at HRS (RESET) the equivalent circuit describing the Nyquist plot is a modified Warburg impedance [22].

### D. Retention measurements

Retention measurements at room temperature denote that RRAM cells on SOI have much smaller resistance loss rate compared to bulk ones.

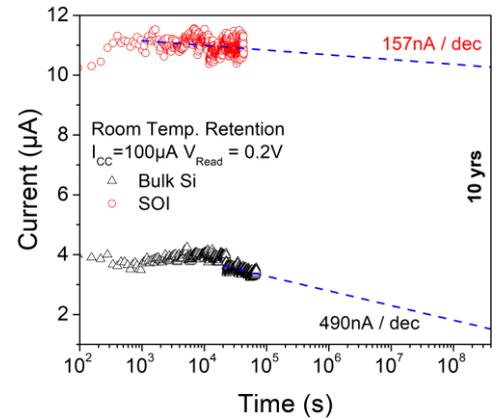

Figure 8. Retentionn measurements at room temperature.

## II. CONCLUSION

The resistance switching memory cells based on silicon nitride insulator fabricated on an SOI wafer exhibit attractive performance characteristics compared with bulk Si wafer in terms of memory window, cell-to-cell variability and retention characteristics.

ACKNOWLEDGMENT

This work was supported in part by the research projects "3D-TOPOS" (MIS 5131411) and "LIMA-chip" (Proj.No. 2748) which are funded by the Operational Program NSRF 2014-2020 and the Hellenic Foundation of Research and Innovation (HFRI) respectively.

REFERENCES

[1] *Beyond CMOS and Emerging Materials Integration,* THE INTERNATIONAL ROADMAP FOR DEVICES AND SYSTEMS (IRDS): 2022, IEEE.
[2] M. Bawedin, S. Cristoloveanu, D. Flandre," Innovating SOI memory devices based on floating-body effects," *Solid-State Electronics,* 51, pp. 1252-1262, 2007, https://doi.org/10.1016/j.sse.2007.06.024.
[3] S. Cristoloveanu, *Fully Depleted Silicon-On-insulator,* Elsevier, 2021, ISBN: 9780128196434
[4] A. Marshall, S. Natarajan, *SOI design: analog, memory, and digital techniques,* Springer New York, NY, 2002
[5] S. V. Suryavanshi, G. Yeric, M. Irby, X. M. H. Huang, G. Rosendale and L. Shifren, "Extreme Temperature (> 200 °C), Radiation Hard (> 1 Mrad), Dense (sub-50 nm CD), Fast (2 ns write pulses), Non-Volatile Memory Technology," *2022 IEEE International Memory Workshop (IMW),* Dresden, Germany, 2022, pp. 1-4, doi: 10.1109/IMW52921.2022.9779251.
[6] Y. Zhou, Z. Cheng, H. Liu, T. Xiong and B. Wang, "A 22-nm FDSOI 8T SRAM Based Time-Domain CIM for Energy-Efficient DNN Accelerators," *2022 IEEE Asia Pacific Conference on Circuits and Systems (APCCAS),* Shenzhen, China, 2022, pp. 501-504, doi: 10.1109/APCCAS55924.2022.10090315
[7] Q. M. Khan, R. Perdriau, M. Ramdani and M. Koohestani, "A Comparative Performance Analysis of 6T & 9T SRAM Integrated Circuits: SOI vs. Bulk," *IEEE Letters on Electromagnetic Compatibility Practice and Applications,* vol. 4, no. 2, pp. 25-30, June 2022, doi: 10.1109/LEMCPA.2022.3163963.
[8] W. Wang, "A Comparative Analyze of FinFET and Bulk MOSFET SRAM Design," *2022 International Conference on Applied Physics and Computing (ICAPC),* Ottawa, ON, Canada, 2022, pp. 211-218, doi: 10.1109/ICAPC57304.2022.00046
[9] N. Mika et al., "Experimental fabrication of an ESF3 floating gate flash cell in an FD-SOI process," *ESSDERC 2022 - IEEE 52nd European Solid-State Device Research Conference (ESSDERC),* Milan, Italy, 2022, pp. 356-359, doi: 10.1109/ESSDERC55479.2022.9947160
[10] Z. Zhou et al., "Inversion-Type Ferroelectric Capacitive Memory and Its 1-Kbit Crossbar Array," *IEEE Transactions on Electron Devices,* vol. 70, no. 4, pp. 1641-1647, April 2023, doi: 10.1109/TED.2023.3243556
[11] P. Dimitrakis, "Introduction to NVM Devices." in *Charge-Trapping Non-Volatile Memories,* P. Dimitrakis, Ed., Springer, Cham, 2015, pp.1-36
[12] J. -C. Liu, C. -W. Hsu, I. -T. Wang and T. -H. Hou, "Categorization of Multilevel-Cell Storage-Class Memory: An RRAM Example," in *IEEE Transactions on Electron Devices,* vol. 62, no. 8, pp. 2510-2516, Aug. 2015, doi: 10.1109/TED.2015.2444663
[13] Q. Xia and J. J. Yang. "Memristive crossbar arrays for brain-inspired computing." *Nature Materials,* vol. 8, no.4, pp. 309-323, 2019
[14] Rao, M., Tang, H., Wu, J. et al. Thousands of conductance levels in memristors integrated on CMOS. *Nature* 615, 823–829 (2023). doi: 10.1038/s41586-023-05759-5
[15] N. Vasileiadis *et al.,* "In-Memory-Computing Realization with a Photodiode/Memristor Based Vision Sensor," *Materials* vol. 14, pp. 5223, 2021
[16] Tan, H., van Dijken, S. "Dynamic machine vision with retinomorphic photomemristor-reservoir computing." *Nat Commun.* 14, 2169 (2023). https://doi.org/10.1038/s41467-023-37886-y

[17] J. -M. Portal et al., "Design and Simulation of a 128 kb Embedded Nonvolatile Memory Based on a Hybrid RRAM (HfO$_2$)/28 nm FDSOI CMOS Technology," *IEEE Transactions on Nanotechnology,* vol. 16, no. 4, pp. 677-686, July 2017, doi: 10.1109/TNANO.2017.
[18] H. Aziza, P. Canet, J. Postel-Pellerin, M. Moreau, J.-M. Portal, and M. Bocquet, "ReRAM ON/OFF resistance ratio degradation due to line resistance combined with device variability in 28 nm FDSOI technology," *Proc. Joint Int. EUROSOI Workshop Int. Conf. Ultimate Integr. Silicon (EUROSOI-ULIS),* Apr. 2017, pp. 35–38, doi:10.1109/ULIS.2017.7962594.
[19] X. Zheng et al., "Back-End-of-Line-Based Resistive RAM in 0.13 μm Partially-Depleted Silicon-on-Insulator Process for Highly Reliable Irradiation- Resistant Application," *IEEE Electron Device Letters,* vol. 42, no. 1, pp. 30-33, Jan. 2021, doi: 10.1109/LED.2020.3037072.
[20] A. Mavropoulis, N. Vasileiadis, C. Theodorou, L. Sygellou, P. Normand, G. Ch. Sirakoulis, P. Dimitrakis, "Effect of SOI substrate on silicon nitride resistance switching using MIS structure," Solid-State Electronics, 194, 108375, (2022), doi: 10.1016/j.sse.2022.108375.
[21] N. Vasileiadis, A. Mavropoulis, P. Loukas, P. Normand, G. C. Sirakoulis and P. Dimitrakis, "Substrate Effect on Low-frequency Noise of synaptic RRAM devices," *2022 IFIP/IEEE 30th International Conference on Very Large Scale Integration (VLSI-SoC),* Patras, Greece, 2022, pp. 1-5, doi: 10.1109/VLSI-SoC54400.2022.9939652
[22] N. Vasileiadis *et al.,* "Understanding the Role of Defects in Silicon Nitride-Based Resistive Switching Memories Through Oxygen Doping," *IEEE Trans. Nanotechnol.* vol. 20, pp. 356–364, 2021.
[23] N. Vasileiadis *et al.,* "Multi-level resistance switching and random telegraph noise analysis of nitride based memristors," *Chaos, Solitons & Fractals* vol. 153, pp. 111533, 2021.
[24] M. Yang, H. Wang, X. Ma, H. Gao, and B. Wang, "Effect of nitrogen accommodation ability of electrodes in SiN$_x$-based resistive switching devices," *Appl. Phys. Lett.,* 111, pp. 233510, 2017.
[25] D.Ielmini and R. Waser, *Resistive Switching: From Fundamentals of Nanoionic Redox Processes to Memristive Device Applications,* Springer, 2016, ISBN: 9783527334179
[26] A. H. Edwards, H. J. Barnaby, K. A. Campbell, M. N. Kozicki, W. Liu and M. J. Marinella, "Reconfigurable Memristive Device Technologies," *Proceedings of the IEEE,* vol. 103, no. 7, pp. 1004-1033, July 2015, doi: 10.1109/JPROC.2015.2441752
[27] Lübben, M., Karakolis, P., Ioannou-Sougleridis, V., Normand, P., Dimitrakis, P. and Valov, I. (2015), "Graphene-Modified Interface Controls Transition from VCM to ECM Switching Modes in Ta/TaO$_x$ Based Memristive Devices." *Adv. Mater.,* 27: 6202-6207. doi:10.1002/adma.201502574
[28] L. Vandelli, A. Padovani, L. Larcher, G. Broglia, G. Ori, M. Montorsi, G. Bersuker, and P. Pavan, "Comprehensive physical modeling of forming and switching operations in HfO$_2$ RRAM devices," *Proc. IEEE IEDM,* 2011, pp. 17.5.1–17.5.4
[29] F. M. Puglisi, L. Larcher, G. Bersuker, A. Padovani and P. Pavan, "An Empirical Model for RRAM Resistance in Low- and High-Resistance States," *IEEE Electron Device Letters,* vol. 34, no. 3, pp. 387-389, March 2013, doi: 10.1109/LED.2013.2238883
[30] K. -I. Takeda, K. Hinode, I. Oodake, N. Oohashi and H. Yamaguchi, "Enhanced dielectric breakdown lifetime of the copper/silicon nitride/silicon dioxide structure," *1998 IEEE International Reliability Physics Symposium Proceedings. 36th Annual* (Cat. No.98CH36173), Reno, NV, USA, 1998, pp. 36-41, doi: 10.1109/RELPHY.1998.670439
[31] K. Okada, Y. Ito and S. Suzuki, "A new model for dielectric breakdown mechanism of silicon nitride metal-insulator-metal structures," *2016 IEEE International Reliability Physics Symposium (IRPS),* Pasadena, CA, USA, 2016, pp. 3A-6-1-3A-6-6, doi: 10.1109/IRPS.2016.7574515